\def\year{2022}\relax
\documentclass[letterpaper]{article} 
\usepackage{aaai22}  
\usepackage{times}  
\usepackage{helvet}  
\usepackage{courier}  
\usepackage[hyphens]{url}  
\usepackage{graphicx} 
\usepackage{amsfonts,amssymb}
\usepackage{amsmath}
\usepackage{bm}
\usepackage{natbib}  
\usepackage{caption} 
\DeclareCaptionStyle{ruled}{labelfont=normalfont,labelsep=colon,strut=off} 
\usepackage{algorithm}
\usepackage{algorithmic}
\usepackage{color}
\usepackage{caption}
\usepackage{subcaption}
\usepackage{tablefootnote}
\usepackage{threeparttable}

\usepackage{newfloat}
\usepackage{listings}
\usepackage{multirow}
\usepackage{epstopdf}
\floatstyle{ruled}
\newfloat{listing}{tb}{lst}{}
\floatname{listing}{Listing}
\title{DemiNet: Dependency-Aware Multi-Interest Network with Self-Supervised Graph Learning for Click-Through Rate Prediction}
\author {
    Yule Wang,\textsuperscript{\rm 1}
    Qiang Luo,\textsuperscript{\rm 2}
    Yue Ding,\textsuperscript{\rm 1}
    Yunzhe Li,\textsuperscript{\rm 1}
    Dong Wang,\textsuperscript{\rm 1}
    Hongbo Deng\textsuperscript{\rm 2}
}
\affiliations {
    \textsuperscript{\rm 1} Shanghai Jiao Tong University, Shanghai, China\\
    \textsuperscript{\rm 2} Alibaba Group, Beijing, China\\
    \{yulewang, dingyue, liyzh28, wangdong\}@sjtu.edu.cn, \{luoqiang.lq, dhb167148\}@alibaba-inc.com\\
}

\usepackage{bibentry}

\begin{document}

\maketitle
\begin{abstract}
Click-through rate (CTR) prediction is  one of the the most important tasks in modern search engine, recommendation and advertising systems.
Recently, some existing models leverage user’s historical behaviors for multiple interest modeling. However, there remain two main challenges in the prior works: (1) Raw user behavior sequence is noisy and intertwined, making it difficult to extract multiple core interests. (2) The latent correlations between extracted multiple interest vectors are neglected, leading to information loss.

In this paper, we propose a novel model named DemiNet (short for \textit{\textbf{DE}pendency-Aware \textbf{M}ulti-\textbf{I}nterest \textbf{Net}work}) to address the above two issues. To be specific, we first consider various dependency types between item nodes and perform dependency-aware heterogeneous attention for denoising and obtaining accurate sequence item representations. 
Secondly, for multiple interests extraction, multi-head attention is conducted on top of the graph embedding. To filter out noisy inter-item correlations and enhance the robustness of extracted interests, self-supervised interest learning is introduced to the above two steps. Thirdly, to aggregate the multiple interests, interest experts corresponding to different interest routes give rating scores respectively, while a specialized network assigns the confidence of each score.
Experimental results on three real-world datasets demonstrate that the proposed DemiNet significantly improves the overall recommendation performance over several state-of-the-art baselines. Further studies verify the efficacy and interpretability benefits brought from the fine-grained user interest modeling.
\end{abstract}



\section{Introduction}
In recommender and advertising systems, the capability of click-through rate (CTR) prediction \cite{wang2021icmt, chen2021airec, shi2022task} not only influences the overall revenue of the whole platform, but also directly affects user experience and satisfaction. 
Recently, most deep CTR models focus on capturing interaction between features from different fields \cite{guo2017deepfm, li2021extracting} and the evolution of user interest over time \cite{zhou2019deep,feng2019deep}. Designing model to capture user’s multiple interests can further improve the performance of CTR prediction as well as the model's interpretability. 

\begin{figure} [t]
  \centering
  \includegraphics[width=0.45\textwidth]{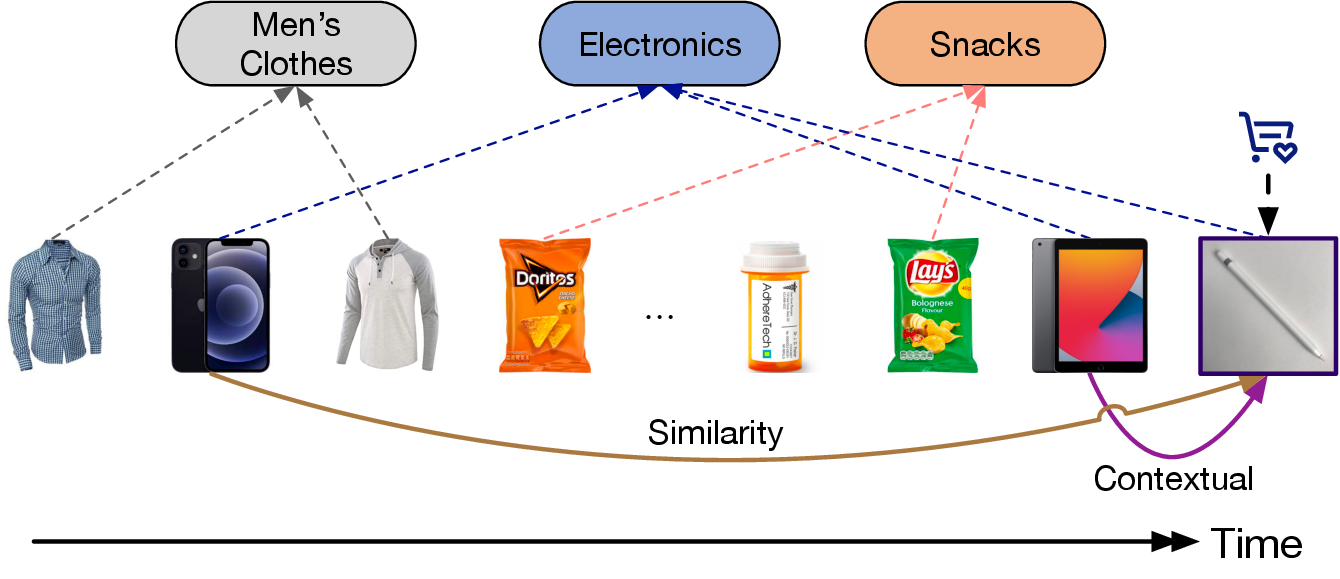}
  \caption{A user usually has multiple interests when browsing an e-commerce platform. His current behavior is highly correlated with short-term contextual and long-term similarity dependencies.}
    \label{multimodal}
\end{figure}

For diverse user interest modeling, Deep Interest Network (DIN) \cite{zhou2018deep} utilizes attention based method to capture relative interests to the target item, and obtains adaptive interest representation. However, compressing diverse user interests into one single vector may lead to information loss. 
Recently, in the matching stage, MIND and ComiRec \cite{li2019multi,cen2020controllable} utilize dynamic routing method based on capsule network \cite{sabour2017dynamic} to adaptively aggregate user’s historical behaviors into multiple representation vectors, which can reflect the various interests of the user. PIMI \cite{chen2021exploring} additionally considers periodicity and graph-structure in the item sequence. DMIN \cite{xiao2020deep} introduces self-attention to capture user's multiple interests in the ranking stage. However, there exist two major challenges in user multiple interest modeling that have not been well-addressed so far as follows:
\begin{itemize}
\item \textbf{User behavior sequences reflect implicit and noisy intention signals, increasing the difficulty of core interest extraction.} The prior methods fully connect all items for better representation learning in the sequence, but actually they introduce a huge amount of noisy interactions. For example, in Figure \ref{multimodal}, currently, the user bought an Apple pencil. Since the user has multiple interests, the actual dependency between this behavior is highly correlated with the short-term contextual dependency and long-term similarity dependency, while its correlation with other historical items is actually weak.
\item \textbf{Given extracted multiple interest vectors, the correlations between them may provide additional information for aggregation, but they were neglected.} Prior methods either utilize the interest vectors separately \cite{chen2021exploring} or just concatenate all the interest vectors \cite{xiao2020deep}, neglecting their underlying relations.
\end{itemize}

To address the above two challenges, we propose a heterogeneous graph enhanced network for robust interest extraction and introduce multiple interest experts for interest aggregation in the ranking stage. Specifically, there are three main components: \textit{Dependency-Aware Interaction Module} serves for denoising and obtaining accurate item representations for the user behavior sequence, in which we novelly construct a heterogeneous graph network with four types of dependencies and perform hierarchical attention.
Later on, as for the \textit{Multi-Interest Extraction Module}, multiple core user interests are extracted through multi-head attention mechanism according to the target item. 
To improve the robustness of the extracted interests, self-supervised interest learning is introduced through performing random edge dropout in the above two modules. 
In the final \textit{Multi-Dependency Interaction Module}, multiple interest experts give prediction ratings respectively. Meanwhile, a Confi-Net calculates the weights of each CTR prediction result. The final output is the weighted aggregation of all the ratings. To summarize, DemiNet makes the following contributions:

\begin{itemize}
\item For denoising and obtaining accurate sequence item representations, we propose a multi-dependency-aware heterogeneous graph attention network to capture the item correlations and then combine the various dependency semantics.
\item We innovatively integrate a self-supervised task into the multi-interest extraction process to filter out noisy correlations between sequence items, enhancing the robustness of the multiple interest representations.
\item We design a novel multiple interests aggregation structure that involves interest experts focusing on corresponding interest routes and a confidence network to aggregate the ratings depending on their interest excitation strengths. 
\item We evaluate our proposed method on three real-world datasets in terms of click-through rate prediction. The results achieve state-of-the-art performance, which verifies the efficacy of DemiNet.
\end{itemize}

\section{Related Work}

\subsection{User Multiple Interest Extraction.} So far, most recent works which explicitly modeling user's diverse interests are based on user behavior sequence. DIN-based methods \cite{zhou2018deep, zhou2019deep} design a local activation unit to adaptively learn the representation of user interests from historical behaviors with respect to the candidate item. 
Despite its functionality, we argue that a unified user embedding is difficult to reflect the user’s multiple interests. On the other hand, MIND \cite{li2019multi} utilizes a dynamic routing method based on the capsule network \cite{sabour2017dynamic} to adaptively aggregate user behavior sequence into multiple representation vectors, which can reflect the different interests of the user. ComiRec \cite{cen2020controllable} first clusters the user behaviors and then leverages self-attention mechanism or dynamic routing method for multi-interest extraction following MIND. PIMI \cite{chen2021exploring} additionally models the periodicity of user’s multiple interests and introduces a graph-structure. DMIN \cite{xiao2020deep} introduces self-attention to refine the behavior sequence embeddings. However, these methods have the following limitations: (1) The interactivity between items in the user behavior sequence is not explored effectively, thus large noisy information is retained in the sequence before interest extraction. (2) During the inference process, the fusion process of multiple interest vectors is coarse-grained and loses semantic information.


\subsection{Graph neural network in CTR Prediction.} Utilizing user-item interaction data, graph neural network(GNN) has achieved great success in multiple large-scale web-based tasks \cite{shi2021versagnn} (i.e., collaborative filtering). In sequential-based CTR prediction scenes, the target of graph representation learning is to refine the embedding of nodes in the graph constructed by the sequence. For example, SRGNN and its follow-ups \cite{wu2019session,xu2019graph} model the session sequence as a directed graph and perform attention mechanism for representation learning. 
LESSR \cite{chen2020handling} further enhances the graph layers to capture more intra-session correlations and maintain the long-range dependencies. SURGE \cite{chang2021sequential} integrates cluster-aware attention in the graph convolutional propagation process. However, the above methods don't fully consider the different types of dependencies in the sequence graph, thus the item-wise interactions are imprecise. 

Heterogeneous graph neural networks (HGNN), on the other hand, can embrace various entities and relationships in a unified graph.  Recent works have introduced HGNN into recommender systems thanks to its heterogeneity and rich semantic information \cite{chen2021graph,jin2020efficient}. HAN\cite{wang2019heterogeneous} utilizes a novel hierarchical attention mechanism to fully consider the particular importance of nodes and meta-paths. However, the underlying dependency relations between item nodes in the sequence graph are not considered in the above works.

\section{Preliminaries}
\subsection{Problem Formalization}
Assume \textit{U}, \textit{I} denotes the user and item set, respectively. Each user $u \in \textit{U}$ has a historical behavior sequence $S^u$ in time order. Historical data collected by the system is collected for building a CTR prediction model. Specifically, each user-item pair instance to be predicted is represented as a tuple (\textit{$S_u$},\textit{$P_u$},\textit{$F_i$}), where $S_u$ represents the user behavior sequence of user $u$, $P_u$ denotes the basic profiles of user $u$ (like user age and gender), $F_i$ denotes the features of target item $i$ (such as item id and category id). 

Given $S^u = (i_1^u,i_2^u,...,i_{n-1}^u,i_{n}^u) $, where $n$ is the number of interactions in the sequence and $i_r^u \in \textit{I}$ represents the $r$-th item in the sequence. In practice, the goal of CTR prediction is to predict the next item $i_{n+1}^u$ that the user might be interested in. 

\subsection{Heterogeneous Graph for Behavior Sequence}

Since the dependency links between sequence items constructed graph is quite comprehensive, heterogeneous graph is a powerful tool which can pack the rich relation semantics into different edge types and graph convolution process. 

For each user behavior sequence $S_u$, we build its heterogeneous directed graph, $\mathcal{G}_u$ = ($\mathcal{V}_u$, $\mathcal{E}_u$). Each vertex $v \in \mathcal{V}_u$ corresponds to an interacted item entity.
To model the comprehensive inter-item dependencies, 
the item-item heterogeneous link interactions are denoted
as $\mathcal{E}_{u} = \{\mathcal{E}_{(1)u}, \mathcal{E}_{(2)u}, ..., \mathcal{E}_{(|R|)u}\}$, where $\mathcal{E}_{(r)u} = [e_{(r)ij}]_{|\mathcal{V}| \times |\mathcal{E}|} \in \{0,1\} $ denotes whether item node $v_i$ has interaction with item node $v_j$ under dependency relation $r$, and $R$ is the set of predefined dependency relation types. 
The input to the heterogeneous graph is a node embedding matrix $\mathbf{H_u} =\{\vec{\mathbf{h}}_1,\vec{\mathbf{h}}_2,\cdots,\vec{\mathbf{h}}_n\}$, $\vec{\mathbf{h}}_i \in \mathbb{R}^d $, where $n$ is the length of the user behavior sequence. Through the graph convolutional network, it produces the refined item node embeddings $\mathbf{H'_u} =\{\vec{\mathbf{h}}_1',\vec{\mathbf{h}}_2',\cdots,\vec{\mathbf{h}}_n'\}$, $\vec{\mathbf{h}}_i' \in \mathbb{R}^d $.

\section{Methodologies}


\subsection{Multi-Dependency Interaction Module}
\begin{figure*} [t]
  \centering
  \includegraphics[width=0.9\textwidth]{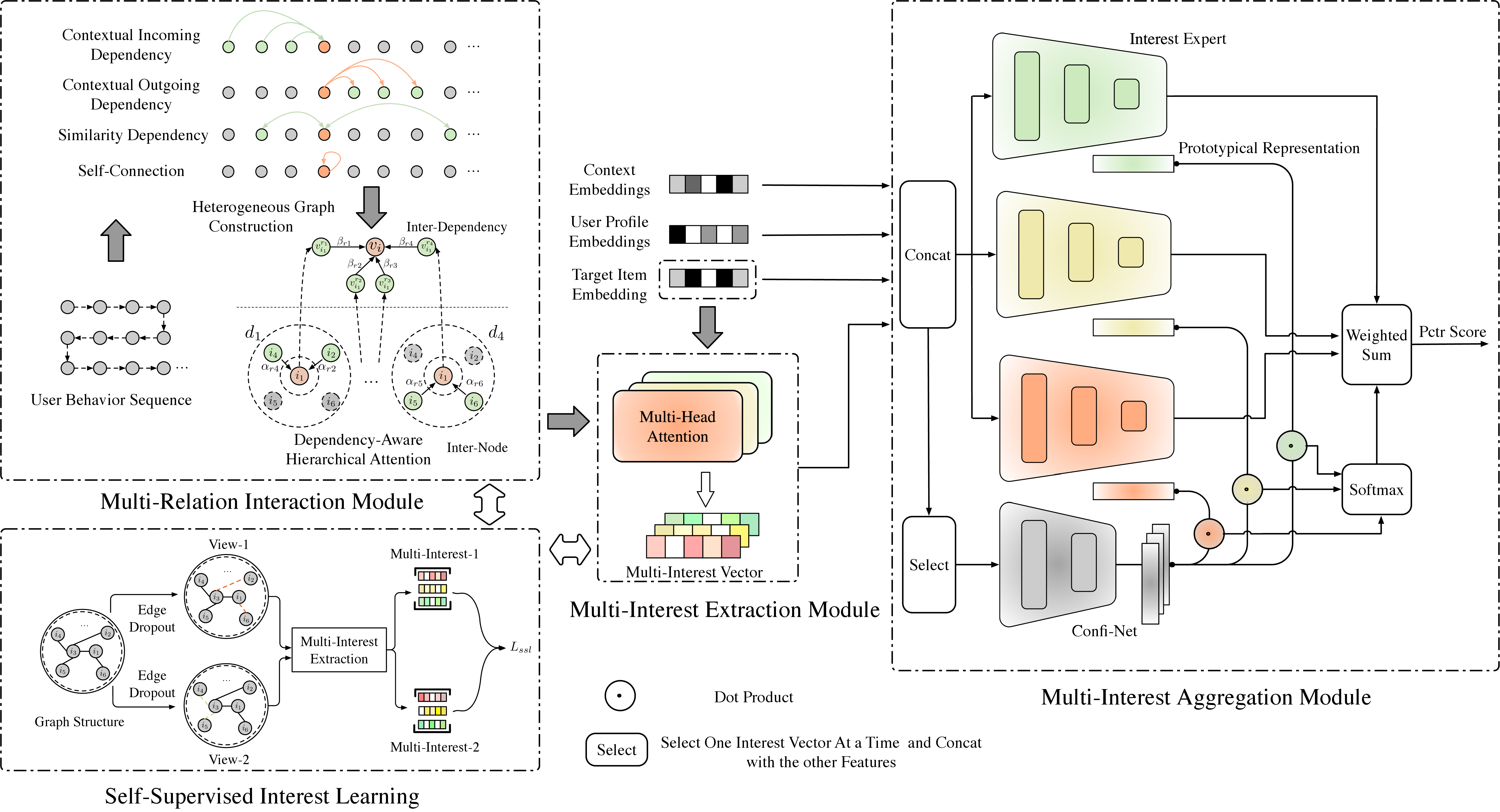}
  \caption{Illustration of the DemiNet model, which is made up of three modules and a self-supervised interest learning task.}
    \label{main_framework}
\end{figure*}

\subsubsection{Heterogeneous Graph Construction}
Specifically, we novelly design the following four inter-node dependency relations in the heterogeneous graph:


\begin{itemize}
\item \textbf{Contextual-Incoming Dependency Relation} \bm{$r_{in}$}. For each node $v_i$ in $\mathcal{G}_u$, the $\epsilon$-prior neighbor set of $v_i$ is denoted as $\mathcal{N}_{\mathcal{\varepsilon}}^{in}(v_i)$, which includes the prior $\epsilon$ items in the behavior sequence. For each item node in $\mathcal{N}_{\mathcal{\varepsilon}}^{in}(v_i)$ , we set it as source node and add a directed edge to $v_i$.
\begin{equation}
\label{input_transition}
e_{(r_{in})ji}= \begin{cases}1, & v_j \in \mathcal{N}_{\mathcal{\varepsilon}}^{in}(v_i); \\ 0, & \text {otherwise; }\end{cases}
\end{equation}
\item \textbf{Contextual-Outgoing Dependency Relation \bm{$r_{out}$}}. For each node $v_i$ in $\mathcal{G}_u$, the $\epsilon$-subsequent neighbor set of $v_i$ is denoted as $\mathcal{N}_{\mathcal{\varepsilon}}^{out}(v_i)$, which includes the subsequent $\epsilon$ items in the behavior sequence. We set $v_i$ as  source node and add a directed edge to each item node in $\mathcal{N}_{\mathcal{\varepsilon}}^{out}(v_i)$.
\begin{equation}
\label{input_transition}
e_{(r_{out})ji}= \begin{cases}1, & v_j \in \mathcal{N}_{\mathcal{\varepsilon}}^{out}(v_i); \\ 0, & \text {otherwise; }\end{cases}
\end{equation}
\item \textbf{Similarity Dependency Relation \bm{$r_{sim}$}}. For every item node pair ($v_i$, $v_j$) in $\mathcal{G}_u$, we perform node similarity measurement and the outcome is recorded as $M_{ij}$. Without loss of generality, we conduct cosine distance \cite{wang2020gcn} as our metric function. If the pair's similarity is larger than the control threshold $t$, we add bidirectional edges between the pair:
\begin{equation}
\label{similarity_metric}
M_{ij} =\cos \left( \vec{\mathbf{h}}_{i}, \vec{\mathbf{h}}_{j}\right).
\end{equation}
\begin{equation}
\label{similarity_dependency}
e_{(r_{sim})ij}, e_{(r_{sim})ji}= \begin{cases}1, & M_{i j} \geq t; \\ 0, & \text {otherwise; }\end{cases}
\end{equation}
\item \textbf{Self-Connection Dependency Relation \bm{$r_{self}$}}. To strengthen the information and uniqueness of itself, we take every item node's self-connection into consideration in $\mathcal{G}_u$.
\begin{equation}
\label{self_connection}
e_{(r_{self})ii}= \begin{cases}1, & \text {$v_i \in \mathcal{V}_u$};  \\ 0, & \text {otherwise; }\end{cases}
\end{equation}
\end{itemize}

Since the scale of each sequence graph $\mathcal{G}_u$ is quite limited, to avoid over-parameterization and over-smoothing \cite{chen2020measuring}, we model the semantics between nodes at the granularity of the above four dependency relations, which are all one-hop. 


\subsubsection{Hierarchical Heterogeneous Graph Attention} To extract core interests and relieve noise signals in $\mathcal{G}_u$, we leverage the graph attention mechanism \cite{velivckovic2017graph} for inter-node interaction since it can strengthen important signals and weaken noise ones during message propagation. Moreover, we notice that neighbors based on different dependency relations play different roles during node interactions. Therefore, each node should firstly fuse message from neighbor nodes of the same dependency and then combine the dependency-wise correlation information.

Specifically, for each node, we first perform inter-node attention to learn the weights of dependency-based neighbor nodes, aggregating them to obtain the dependency-specific representation. After that, inter-dependency attention is performed to obtain the unified node embedding, which is the adaptive weighted combination of dependency-specific representations. To increase the expressive power and stabilize the learning process, the attention process is extended to the multi-head metric \cite{vaswani2017attention}.
\begin{itemize}
\item \textbf{Inter-Node Attention.} Under the dependency-specific edges $\mathcal{E}_{(r)}$, given a node embedding pair of head $\phi$ $(\vec{\mathbf{h}}_{i}^\phi, \vec{\mathbf{h}}_{j}^\phi)$ with directed connection, the inter-node attention can learn the importance $a_{(r)ij}^\phi$ which means how important node $v_j$ is for node $v_i$:
\begin{equation}
\label{node_importance}
a_{(r)ij}^{\phi} =  \text{LeakyReLU} (\mathbf{W_n^{(r)\phi}} \otimes [ \vec{\mathbf{h}}_{i}^\phi||\vec{\mathbf{h}}_{j}^\phi]),
\end{equation}
where $\mathbf{W_n^{(r)\phi}}$ is a trainable matrix, $\otimes$ is the concatenation and matrix product operator. Then we normalize the importance to get the weight coefficient $\alpha_{(r)ij}^\phi$ via softmax function:
\begin{equation}
\label{node_weights}
\alpha_{(r)ij}^\phi = \frac{\exp (a_{(r)ij}^\phi)}{\sum_{k \in \mathcal{N}_{i}^{(r)}} \exp (a_{(r)ik}^\phi)}.
\end{equation}
The node embedding of $v_i$ under relation $r$ can be aggregated as follows, in which $\|$ is the concatenation operator:
\begin{equation}
\label{node_aggregate}
\vec{\mathbf{h}}_{(r)i} = \bigg|\bigg|_{\phi=1}^{\Phi} \sum_{k \in \mathcal{N}_{i}^{(r)}} \alpha_{(r)ik}^\phi \cdot \vec{\mathbf{h}}_{k}^\phi.
\end{equation}

\item \textbf{Inter-Dependency Attention.} For fusing semantic information of node $v_i$, after obtaining all the dependency-specific embeddings, dependency-level attention first learns the importance of each dependency embedding, denoted as $a_{(r)i}^d$, is shown as follows:
\begin{equation}
\label{relation_importance}
e_{(r)i}^\phi = \text{LeakyReLU} (\mathbf{W_d^{\phi}} \otimes \vec{\mathbf{h}}_{(r)i}^{\phi} + b_d^{\phi}),
\end{equation}
where $\mathbf{W_d^{\phi}}$ is a trainable matrix, $b_d^{\phi}$ is a trainable scalar, $\otimes$ is the matrix product operator. The weight of dependency relation $r$ of node $v_i$, denoted as $\beta_{(r)i}^\phi$, can be obtained by normalizing the importances of all dependency relations,
\begin{equation}
\label{relation_weights}
\beta_{(r)i}^{\phi} = \frac{\exp (e_{(r)i}^{\phi})}{\sum_{k \in R} \exp (e_{(k)i}^{\phi})}.
\end{equation}
Then, the unified embedding of node $v_i$ under dependency relation $r$ can be aggregated as follows, in which $\|$ is the concatenation operator::
\begin{equation}
\label{relation_aggregate}
\vec{\mathbf{h}}_{i}^{\prime}= \|_{\phi=1}^{\Phi} \sum_{k \in R} \beta_{(k)i} \cdot \vec{\mathbf{h}}_{(k)i}^{\phi}.
\end{equation}

\end{itemize}


After obtaining refined item embeddings $\mathbf{H'_u} \in \mathbb{R}^{n \times d}$ from the prior module, to strengthen the sequence order relationship, we add it with a trainable positional embedding matrix $\mathbf{E_p} \in \mathbb{R}^{n \times d}$, the final node embeddings $\mathbf{H^*_u} \in \mathbb{R}^{n \times d}$ is shown as follows:
\begin{equation}
\label{final_node_embdding}
\mathbf{H^*_u = H'_u + E_p}.
\end{equation}

\subsection{Multi-Interest Extraction Module}
Since multi-interest extraction can be viewed as the process of soft clustering of items, we utilize multi-head attention \cite{vaswani2017attention} to extract multiple interests, in which each head captures a unique interest vector. 
To be specific, given the candidate item embedding $\vec{\mathbf{h}}_t$ and item sequence embedding $\mathbf{H^*_u}$, the interest embedding $\overrightarrow{\mathbf{v}}_{ui}^k$ is generated as follows:
\begin{equation}
\label{v_ui}
\overrightarrow{\mathbf{v}}_{ui}^k = \sum_{i=1}^{n} \mathbf{Attention_{k}}(\vec{\mathbf{h}}_{i}^*||\vec{\mathbf{h}}_{t}) \cdot \vec{\mathbf{h}}_{i}^*
\end{equation}
where the attention mechanism $\mathbf{Attention_{k}}$ is a two-layer feed-forward neural network applying the LeakyReLU nonlinearity, $\|$ is the concatenation operator.
Through stacking all the captured interest vector representations from $K$ attention heads, we can obtain the user multiple interests matrix according to the target item $i$: $\mathbf{V}_{ui} = \{\overrightarrow{\mathbf{v}}_{ui}^1,\overrightarrow{\mathbf{v}}_{ui}^2,\cdots,\overrightarrow{\mathbf{v}}_{ui}^K\} \in \mathbb{R}^{K\times d}$.


\subsection{Multi-Interest Aggregation Module}
Obtaining $\mathbf{V}_{ui}$ from the prior module, in this place we design $K$ interest experts focusing on various domains to give prediction scores respectively. Meanwhile, the confidence level of each score is calculated according to a specific Confi-Net which is based on a three-layer neural network. The final output of the module is the weighted aggregation of the prediction scores.
\subsubsection{Interest Expert} This unit contains the main CTR prediction route and a trainable interest expert prototypical representation $\mathbf{p_k}$. As for the main CTR prediction route, it contains a unique batch normalization layer following by a three-layer neural network. Given $\mathbf{V}_{ui}$, we flatten it and concatenate them with embeddings of $S_u$ and $F_i$, denoting this overall embedding vector as $\mathbf{E}_{ui}^{\prime}$. With $\mathbf{E}_{ui}^{\prime}$ as the network input, through the main route, each interest expert $k$ produces its output score $o_{ui}^k$.

\subsubsection{Confi-Net} The implementation of Confi-Net is a three-layer MLP. Given the $k$-th interest expert, we fetch the $k$-th user interest vector $\overrightarrow{\mathbf{v}}_{ui}^k$ from $\mathbf{V}_{ui}$
and concatenate it with $S_u$ and $F_i$, denoting this embedding vector as $\mathbf{E}_{ui}^k$. Feeding $\mathbf{E}_{ui}^k$ into the Confi-Net, it outputs a combination vector $\mathbf{c}_{ui}^k$, which condenses the excitation relationship between the candidate item and the $k$-th interest. We perform dot product to calculate the confidence between $\mathbf{c}_{ui}^k$ and the $k$-th interest expert. The final confidence of each interest expert is assigned with a normalized weighted $\omega_{ui}^k$:
\begin{equation}
\label{omege_calculation}
\omega_{ui}^k = \frac{ \exp (\frac{1}{\sqrt{d}} \mathbf{c}_{ui}^k \cdot \mathbf{p}_k)}{\sum_{k=1}^K \exp (\frac{1}{\sqrt{d}} \mathbf{c}_{ui}^k \cdot \mathbf{p}_k)}.
\end{equation}
We have scaled the dot product results by a factor of $\frac{1}{\sqrt{d}}$ to help convergence. Then we get final output vector of the model $\hat{y}_{ui}$:
\begin{equation}
\label{y_output}
\hat{y}_{ui} = \operatorname{softmax} (\sum_{k=1}^K \omega_{ui}^k \cdot o_{ui}^k)
\end{equation}
It should be noticed that in Eq.\ref{omege_calculation}, during the softmax process, the model is already forced to preserve distinct information in each interest route \cite {ma2020disentangled}. Therefore, in practice, we do not add regularization terms to encourage disentanglement between the $K$-routes interest vectors.
\subsection{Self-Supervised Interest Learning}
The heterogeneous graph network empowers DemiNet to acquire accurate item sequence representations. However, after modeling the four types of dependency relations, we consider that the denseness of inter-item correlations in sequence graph might impede interest extraction process, which would result in a sub-optimal CTR prediction performance. Inspired by the successful practices of self-supervised learning on simple graphs \cite{wu2021self}, we novelly integrate it into our model to enhance the robustness of interests extraction process. We first perform random edge dropout on the heterogeneous graph twice, generating two different graph views. Then by graph attention and interest extraction, we can obtain two views of multiple interests. Through Jensen–Shannon(JS)-Divergence minimization between the two interest views, the attention operators and interest extractor can capture more robust features and relieve the impact from noisy inter-item correlations.
\subsubsection{Multi-View Generation}
For the heterogeneous graph $\mathcal{G}_u$, we perform edge dropout with the dropout ratio $\rho$, creating two different views of $\mathcal{G}_u$. The two independent edge dropout processes $s_{1}$ and $s_{2}$ are represented as:
\begin{equation}
\label{ED_1}
s_{1}(\mathcal{G}_u)=\left(\mathcal{V}_u, \mathbf{M}' \odot \mathcal{E}_u\right),
\end{equation}
\begin{equation}
\label{ED_2}
s_{2}(\mathcal{G}_u)=\left(\mathcal{V}_u, \mathbf{M}'' \odot \mathcal{E}_u\right).
\end{equation}
 where $\mathbf{M}', \mathbf{M}'' \in\{0,1\}^{|\mathcal{E}_u|}$ are two masking vectors on the edge set $\mathcal{E}_u$. Then we go through the graph attention and interest extraction forward pass on $s_{1}(\mathcal{G}_u)$ and $s_{2}(\mathcal{G}_u)$ respectively, obtaining two multiple interests matrix $\mathbf{V}_{ui}^{\prime}, \mathbf{V}_{ui}^{\prime \prime}$ from different views. Since in the heterogeneous graph construction process, not all the edges in the dependency relations are valuable, some redundant edges are introduced. Hence, for denoising, we couple these two interest matrixs together, aiming to capture the intrinsic and core patterns of the dependency relations. \\

\subsubsection{Self-Supervised Regularization}
Specifically, we apply softmax along each interest vector in the two interests matrixs, obtaining their dimension-level interest distributions. For $\mathbf{V}_{ui}^{'}, \mathbf{V}_{ui}^{''}$, the two distributions at interest-route $k$ are denoted as $\mathcal{P}_{(k)ui}^{'}, \mathcal{P}_{(k)ui}^{''}$. Then, we conduct regularization by minimizing the JS divergence between these two output distributions. After summing up all the training samples, the regularization term is formulated as:
\begin{equation}
\begin{aligned}
\label{JSD}
\mathcal{L}_{ssl}^{k}= & \frac{1}{2} \sum_{u \in U} \sum_{i \in I} \mathcal{D}_{KL}\left(\mathcal{P}_{(k)ui}^{'}|| \mathcal{P}_{(k)ui}^{''}\right) \\ + &
\frac{1}{2} \sum_{u \in U} \sum_{i \in I} \mathcal{D}_{KL}\left(\mathcal{P}_{(k)ui}^{''}|| \mathcal{P}_{(k)ui}^{'}\right)
\end{aligned}
\end{equation}
\subsubsection{Overall Loss Function}
The main loss function is defined as the cross-entropy of the prediction and the ground truth. It can be written as follows:
\begin{equation}
\begin{aligned}
\label{CE}
\mathcal{L}_{ce} =- \sum_{u \in U} \sum_{i \in I} ( & y_{ui} \log \left(\hat{y}_{ui}\right) \\ + & \left(1-y_{ui}\right) \log \left(1-\hat{y}_{ui}\right))
\end{aligned}
\end{equation}
where $y_{ui}$ denotes the one-hot encoding vector of ground truth.
Considering the self-supervised regularization, the final training function of DemiNet is formulated as:
\begin{equation}
\label{overall_loss}
\mathcal{L} = \mathcal{L}_{ce} + \beta \sum_{k=1}^{K} \mathcal{L}_{ssl}^{k}
\end{equation}
where $\beta$ controls the magnitude of the self-supervised task.

\section{Experiments}


\subsection{Experimental Setup} 

\subsubsection{Datasets} We conduct experiments on three publicly available datasets: Taobao\footnote{https://tianchi.aliyun.com/dataset/dataDetail?dataId=649\&use\\rId=1}, Amazon\footnote{http://jmcauley.ucsd.edu/data/amazon/} and RetailRocket\footnote{https://www.kaggle.com/retailrocket/ecommerce-dataset}. Taobao dataset is a collection of user behaviors from Taobao’s recommender system. It contains user behavior sequences of about one million users. We take the click behaviors for each user and sort them according to time for constructing the behavior sequence. For the Amazon Dataset, we choose the \textit{Books} subset. We regard reviews as one kind of interaction behavior, and sort the reviews from one user by time. RetailRocket is collected from a real-world e-commerce website. It contains sequential events of viewing and adding to cart. We treat both of them as clicks. After preprocessing, the statistics of the datasets are shown in Table \ref{datasetstatistics}.

\begin{table}
\begin{center}
\caption{Dataset Statistics.}
    \label{datasetstatistics}
\resizebox{\linewidth}{!}{
\begin{tabular}{lcccc}
\hline Dataset & Users & Items & Categories & Samples \\
\hline Taobao & 987,994 & 4,162,024 & 9,439 & 11,526,010 \\
Amazon(Book) & 603,668 & 367,982 & 1,602 & 4,736,082 \\
RetailRocket & 422,274 & 417,053 & 1,669 & 4,926,539 \\
\hline
\end{tabular}}
\end{center}
\end{table}

\subsubsection{Settings and Evaluation Metrics}To make sure the strong generalization of our method and avoid data leakage, we split the train set and test set rigidly according to the timestamp. Specifically, we search the $80\%$ time splitting point of each dataset, putting behavior samples occurred before that splitting point into train set while latters into the test set. For all the datasets, we filter out users with less than 5 interactions. For evaluation metrics, we choose the widely used AUC and Log Loss to measure the performance. In the CTR prediction task, they reflect the pairwise ranking ability and point-wise likelihood, respectively.

\subsubsection{Baselines} To illustrate the effectiveness of our model, we choose methods in the following two groups for comparison:
\begin{itemize}
\item \textbf{Group 1 (Multi-Interest Extraction)} (1) \textbf{DIN} \cite{zhou2018deep} is an early work for user behavior modeling, which proposes to soft-search user behaviors w.r.t. candidates. (2) \textbf{DIEN} \cite{zhou2019deep} integrates GRU with attention mechanism for capturing the evolution trend of user interests. We omit the trick of auxiliary loss for better embedding learning. (3) \textbf{MIND} \cite{li2019multi} extracts multi-interest through dynamic routing. (4) \textbf{ComiRec} \cite{cen2020controllable} combines multi-interest extraction and diversity-aware regularization together. There exists two versions: ComiRec-SA and ComiRec-DR, which based on attention mechanism and dynamic routing respectively. (5) \textbf{DMIN} \cite{xiao2020deep} captures user's latent multiple interests through stacking multi-head attention layers in the ranking stage.
\item \textbf{Group 2 (Ensemble Learning)} (1) \textbf{Multi-Avg} gives the average score of all the interest experts. (2) \textbf{Hard-Routing}  chooses the score from the interest expert with the highest confidence as the final result in the inference stage. (3) \textbf{MoE} \cite{ma2018modeling} designs a mixture of experts, in which every expert's probability is obtained through a gating network.
\end{itemize}

\subsubsection{Parameter Settings}
All methods are learned with the Adam optimizer \cite{kingma2014adam}. The batch size is set as 256. The learning rate is set as $10^{-3}$ . For a fair comparison, the embedding size of each feature $d$ is set as 16 for all models. For the self-supervised interest learning part, the edge-dropout rate is set as 0.6 on all datasets. For hyperparameters of DemiNet, the similarity threshold $t$ is searched between \{0.6, 0.7, 0.8\}. the number of neighbors $\mathcal{\varepsilon}$ in contextual dependency modeling is set as $3$. We set $K = 8$ on the Taobao dataset while $K = 4$ on the other two datasets according to the gap statistic method \cite{tibshirani2001estimating}. To avoid over-smoothing on the dense heterogeneous graph, we set the depth of GNN layers as $2$.

\subsection{Overall Performance Comparison }
The overall experimental results are shown in Table \ref{big_table}. 
\begin{itemize}
\item \textit{Group 1 (Multi-Interest Extraction)} From the results, we could find the following facts: (i) The performance of DemiNet is significantly better than the baselines. AUC values are improved by 3.25\%, 1.66\%, and 1.76\% on three datasets compared to the best baseline, respectively. (ii) We could find that multi-vector interest representation models (DMIN, ComiRec\_SA) outperform other traditional models which only use one unified user interest embedding. This fact shows that extracting multiple interests is essential to CTR prediction performance. (iii) DemiNet outperforms other multiple interest extraction methods significantly. MIND \& ComiRec neglect the interactions in item sequences, failing to denoise it. Moreover, the interests are relatively independent in their aggregation part. DMIN on the other hand introduces noisy item interactions.
\item \textit{Group 2 (Ensemble Learning)} In this section, we modify the aggregation method of scores from interest experts. From the table, we can observe that: (i) DemiNet outperforms all the baselines in three datasets with 0.93\%, 0.79\%, and 0.34\% improvement rates, respectively. (ii) Multi-Avg and Hard-Routing are two simple yet effective aggregation operators. DemiNet outperforms them, manifesting the importance of each interest expert is adaptive according to the user-item pair. (iii) DemiNet's improvement over MoE manifests that obtaining the excitation relation between user interest and candidate item before interacting with the interest expert representation is an effective approach, which is a clear two-step process.

\end{itemize}

\begin{table}
\begin{center}
\caption{The performance of DemiNet with other baseline methods over three datasets}
\label{big_table}
\resizebox{\linewidth}{!}{
\begin{tabular}{ccccccccc}
\hline 
\multirow{2}{*}{ Method } & \multicolumn{2}{c} { Taobao } & \multicolumn{2}{c} { Amazon(Book) } & \multicolumn{2}{c} { RetailRocket } \\
\cline { 2 - 3 } \cline { 4 - 5 } \cline { 6 - 7 } & AUC & LL & AUC & LL & AUC & LL \\
\hline 
DIN & $0.8731$ & $0.2235$ & $0.7469$ & $0.2912$ & $0.6838$ & $0.3212$ \\
DIEN & $0.8738$ & $0.2237$ & $0.7503$ & $0.2902$ & $0.6842$ & $0.3177$ \\
MIND & $0.8695$ & $0.2228$ & $0.7453$ & $0.2921$ & $0.6925$ & $0.3160$ \\
ComiRec\_DR & $0.8696$ & $0.2241$ & $0.7429$ & $0.2923$ & $0.6832$ & $0.3224$ \\
ComiRec\_SA & $0.8768$ & $0.2186$ & $0.7517$ & $0.2891$ & $0.6937$ & $0.3150$ \\
DMIN & $\underline{0.8852}$ & $\underline{0.2171}$ & $\underline{0.7537}$ & $\underline{0.2881}$ & $\underline{0.6967}$ & $\underline{0.3143}$ \\
\hline
Hard-Routing & $0.8953$ & $0.2133$ & $0.7568$ & $0.2873$ & $0.6990$ & $0.3141$ \\
Multi-Avg & $0.8987$ & $0.2125$ & $0.7590$ & $0.2868$ & $0.7018$ & $0.3128$ \\
MoE & $\underline{0.9010}$ & $\underline{0.2118}$ & $\underline{0.7612}$ & $\underline{0.2857}$ & $\underline{0.7042}$ & $\underline{0.3110}$ \\
\hline
\textbf{DemiNet} & $\mathbf{0.9053^*}$ & $\mathbf{0.2045^*}$ & $\mathbf{0.7642^*}$ & $\mathbf{0.2830^*}$ & $\mathbf{0.7059^*}$ & $\mathbf{0.3096^*}$ \\
\hline
\end{tabular}}
  \begin{tablenotes}\footnotesize
    \item $*$ denotes significance p-value \textless 0.01 versus best baseline.
\end{tablenotes}
\end{center}
\end{table}

\begin{table}
\begin{center}
\caption{Ablation Study}
\label{ablation_study}
\resizebox{\linewidth}{!}{
\begin{tabular}{ccccccccc}
\hline 
\multirow{2}{*}{ Method } & \multicolumn{2}{c} { Taobao } & \multicolumn{2}{c} { Amazon(Book) } & \multicolumn{2}{c} { RetailRocket } \\
\cline { 2 - 3 } \cline { 4 - 5 } \cline { 6 - 7 } & AUC & LL & AUC & LL & AUC & LL \\
\hline
w/o DHA & $0.8925$ & $0.2162$ & $0.7551$ & $0.2890$ & $0.6979$ & $0.3167$ \\
w/o SSL& $0.9040$ & $0.2049$ & $0.7633$ & $0.2841$ & $0.7049$ & $0.3107$ \\
w/o IE & $0.8943$ & $0.2158$ & $0.7561$ & $0.2880$ & $0.6987$ & $0.3154$ \\
\hline
\textbf{DemiNet} & $\mathbf{0.9053^*}$ & $\mathbf{0.2045^*}$ & $\mathbf{0.7642^*}$ & $\mathbf{0.2830^*}$ & $\mathbf{0.7059^*}$ & $\mathbf{0.3096^*}$ \\
\hline
\end{tabular}}
  \begin{tablenotes}\footnotesize
    \item $*$ denotes significance p-value \textless 0.01 versus best baseline.
\end{tablenotes}
\end{center}
\end{table}

\begin{figure} [t]
     \centering
     \begin{subfigure}[b]{0.225\textwidth}
         \centering
         \includegraphics[width=\textwidth]{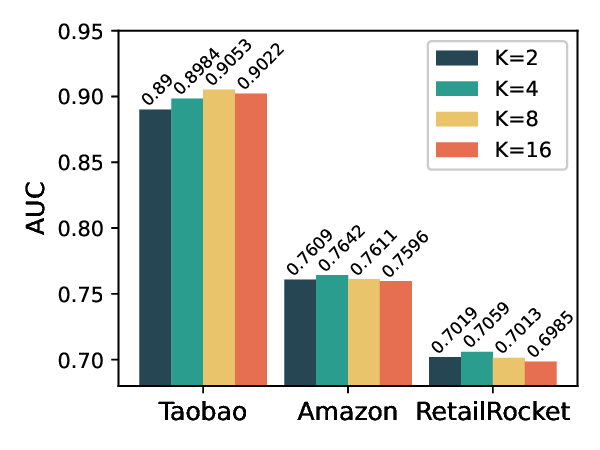}
         \caption{Interest Route Number $K$}
         \label{hyper_K}
     \end{subfigure}
     \hfill
     \begin{subfigure}[b]{0.225\textwidth}
         \centering
         \includegraphics[width=\textwidth]{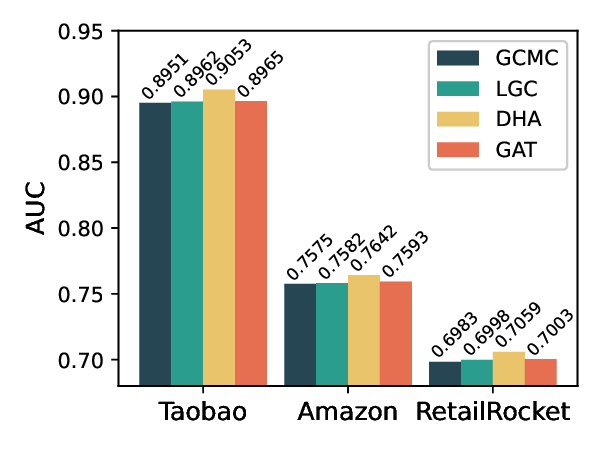}
         \caption{GCN Operator}
         \label{hyper_gcn}
     \end{subfigure}
     \caption{Hyper-Parameter Study}
    \label{hyper_study}
\end{figure}

\begin{figure} [t]
     \centering
     \begin{subfigure}[b]{0.225\textwidth}
         \centering
         \includegraphics[width=\textwidth]{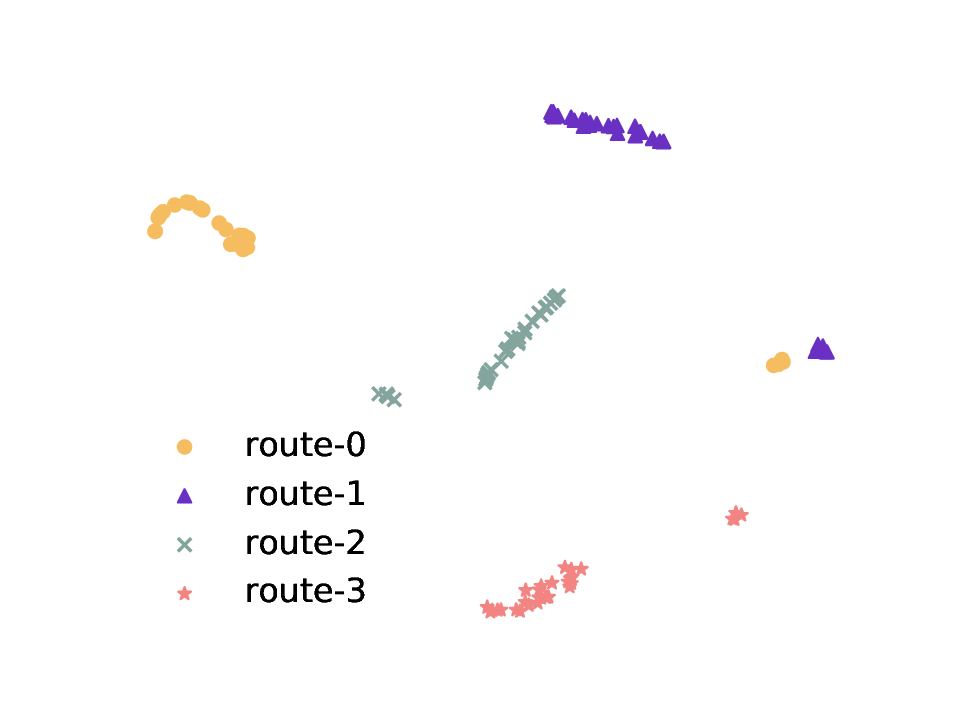}
         \caption{$u_{62226}$}
         \label{cs_scatter_1}
     \end{subfigure}
     \hfill
     \begin{subfigure}[b]{0.225\textwidth}
         \centering
         \includegraphics[width=\textwidth]{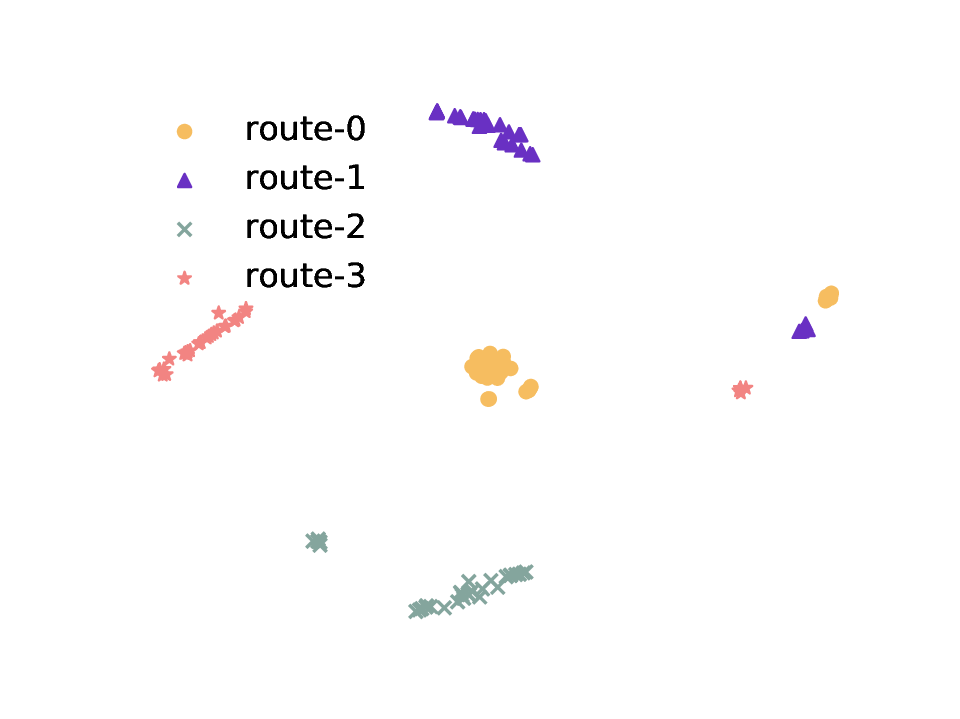}
         \caption{$u_{87539}$}
         \label{cs_scatter_2}
     \end{subfigure}
     \caption{Multiple Interests Visualization using t-SNE }
    \label{case_study_scatter}
\end{figure}

\begin{figure} [t]
  \includegraphics[width=0.45\textwidth]{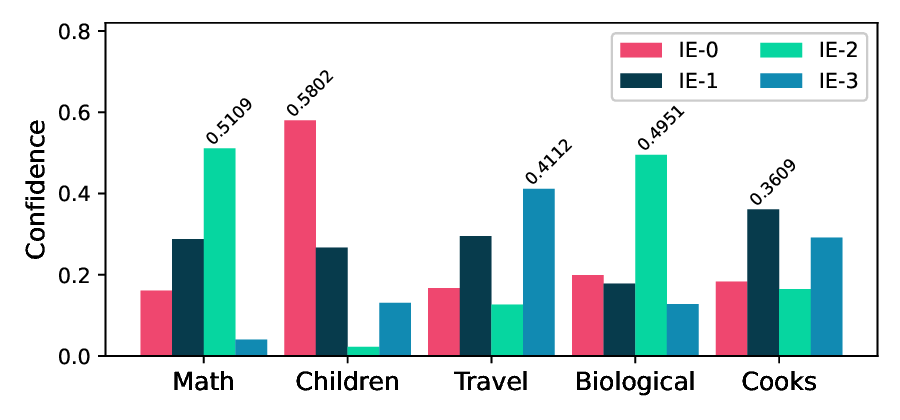}
  \caption{Case Study Of the Assigned Confidence}
    \label{case_study_weights}
\end{figure}

\subsection{Ablation and Hyper-Parameter Study}
\subsubsection{Ablation Study}We further investigate that multi-dependency-aware heterogeneous attention (DHA), self-supervised interest learning and multi-interest aggregation module are all essential parts of the CTR predcition task. We conduct the ablation studies by removing the above three components. For the model variant w/o DHA, we remove the heterogeneous graph neural network and extract the multiple interests directly from the initial behavior sequence. And for the model variant w/o SSL, we remove the self-supervised interest learning task. In w/o IE, we replace multiple interest experts with a single one. We show the experimental results of them on the three datasets in Table \ref{ablation_study}. According to the experimental results, we have the following observations:
\begin{itemize}
\item DemiNet performs better than w/o DHA, w/o SSL and w/o IE in terms of AUC and Log Loss, which demonstrates each component improves the performance effectively.
\item Among the three main components, DHA brings the most significant performance gain through modeling the contextual and similarity item dependencies. Multiple interest experts also enhances the prediction accuracy greatly through adaptive aggregation. As for self-supervised interest learning, we can draw to the conclusion that more robust interest representations can improve the performance by extracting core interests.
\end{itemize}

\subsubsection{Effect of Multiple Interest Number $K$} Figure \ref{hyper_K} shows the performance comparison for the number of interests $K$ on all the three datasets, in which we vary $K$ from $2$ to $16$. The experimental results demonstrate that as the number of interests increases, the model captures and combines multiple intentions of the user and its performance will be higher. However, the performance tends to become saturated as $K$ continues to grow, possibly due to overfitting problem, and the computational cost will also increase. 

\subsubsection{Effect of Graph Convolution Operator} We compare the effects of using various graph convolutional operators on the heterogeneous graph, including GCMC \cite{berg2017graph}, LGC \cite{he2020lightgcn} and vanilla GAT \cite{velivckovic2017graph} and our proposed DHA. The first observation is that methods based on attention mechanism performs slighter better, which can strengthen important signals and weaken noisy signals. The second observation is that DHA performs the best on all the datasets, which demonstrates the information gain of considering the dependency semantics into graph attentive process.


\subsection{Case Study }

\subsubsection{In-depth analyses of the extracted multiple interests}
For in-depth analyzing the extracted user multiple interests, we randomly select two users in Amazon-Book dataset and project their interests vectors according to various candidate items into 2-D space with t-SNE algorithm \cite{van2008visualizing}. As shown in Figure \ref{case_study_scatter}, dots with the same color refers to the same interest route. From the distribution we can observe that interest vectors from the same fine-grained interest route tend to cluster with each other. To conclude, the visualization shows DemiNet can effectively distinguish the fine-grained implicit intentions behind the interactions between users and items, leading to finer reflection of deep motivations behind user behaviors.
\subsubsection{Adaptive confidence assignment of various items}
To analyze the confidence assignment variation according to the candidate items, we randomly select a user in Amazon-Book dataset. Figure \ref{case_study_weights} shows five recommended items for this user which belong to different categories.

As for items belonging to similar category in semantics (i.e. Math and Biological in this case), DemiNet tends to assign higher confidence to one same interest expert (i.e. IE-2 in this case), which is skilled in that domain. While for items belonging to categories that are quite distant in semantics, the interest expert assigned with higher confidence is also different.
\section{Conclusion}
In this paper, we propose a novel method named DemiNet which models the multiple user interests in CTR prediction tasks explicitly. To alleviate the noisy signals in the behavior sequence, we perform multi-dependency-aware heterogeneous attention and self-supervised interest learning. To aggregate the multiple interests, route-specific interest experts and Confi-Net are introduced. We conduct extensive experiments on three real-world datasets. The results demonstrate the effectiveness of DemiNet. Future work includes multi-hop dependency meta-path modeling and investigating the interpretability of DemiNet.

\bibliography{HEAM.bib}

\end{document}